\documentclass[twocolumn,showpacs,preprintnumbers,amsmath,amssymb]{revtex4}
\usepackage{mathrsfs}

\usepackage{graphicx}
\usepackage{dcolumn}
\usepackage{bm}

\begin{document}

\preprint{}

\title{Casimir Forces in Trapped Dilute Bose Gas Between Two Slabs }

\author{Xiao-Lu Yu$^{1,2}$, Ran Qi$^{1}$, Z.B. Li$^{2}$ and W.M. Liu$^{1}$}
\address{$^1$Beijing National Laboratory for Condensed Matter Physics,
Institute of Physics, Chinese Academy of Sciences, Beijing 100080,
China}
\address{$^2$The State Key Laboratory of Optoelectronic Materials and Technologies, School of Physics and Engineering, Sun Yat-Sen University, Guangzhou 510275, China}

\date{\today}

\begin{abstract}
We discuss a Casimir force due to zero-temperature quantum
fluctuations in a weakly interacting Bose-Einstein condensate (BEC)
with a strong harmonic trap. The results show that the presence of a
strong harmonic trap changes the power law behavior of Casimir force
due to the dimensional reduction effect. At finite temperature, we
calculate Casimir force due to thermal fluctuation and find an
exotic temperature dependent behavior in the first order term.
Finally, we speculate some possible experimental realization and
detection of the force in future experiments.
\end{abstract}

\pacs{05.30.Jp, 03.75.Kk, 31.30.Jv, 67.57.De}
\maketitle

The original Casimir effect indicates an attractive force between
two ideal conducting plates in the vacuum due to the fluctuation of
the ground state of quantum electrodynamic. At zero temperature,
there are no real photons in between two neutral plates. So this
force is a pure quantum effect which comes from the remarkable
properties of quantum field theory. This effect has been
experimentally confirmed in 1997 \cite{Lamoreaux} and 1998
\cite{Mohideen}. An analogous effect can be found in a confined BEC
system due to the quantum fluctuation of the ground state at zero
temperature or the thermal fluctuation at finite temperature. The
reason why we are interested in this effect is that it is the first
time we can realize the Casimir effects in a quantum matter system.
In this case, the vacuum fluctuation of the electromagnetic field is
analogous to fluctuation of the ground state of the BEC system,
which is the quasi-particle (Bogoliubov phonon). With the rapid
developments of experimental technics in cold atom system, the
measurement of the Casimir effects in this system will be more
controllable to match various configurations of theoretical
proposals.

Recently, the Casimir effects in BEC system have attracted a lot of
interest both in experiments and theories. Indirect experimental
effects has been observed \cite{Stringari,Kurn,Greiner,Vogels}.
Theoretically, the Casimir force in several models of BEC system has
been studyed. The Casimir effect of three-dimensional ideal bose gas
has been considered at finite temperature with or without traps
\cite{Shyamal1,Oshmyansky,Martin,Antezza}. The interaction between
the atoms has been taken into account at zero temperature without
trap \cite{Roberts,Edery}. For the first time, we consider the
effect of trapping potential and the atom-atom interaction
simultaneously which will both contributes to the Casimir force in
this paper.

The purpose of the present paper is to investigate the behavior of
Casimir force of BEC system in the case of strong harmonic trap. The
geometry of the trapping plays a crucial role in both experimental
and theoretical aspects. In experiments, the presence of trapping
not only make BEC system more realistic, but also provide a new way
to control the Casimir force compared with theoretical predictions.
Theoretically speaking, with the help of strong harmonic trap, we
will open a new window to understand more about dimensional
reduction effect.

In this paper, we will first study the Casimir effect of weakly
interacting dilute Bose gas between two slabs at zero temperature.
With a strong harmonic trap to make system more experimentally
controllable, we modify the Bose gas into quasi-one dimension
\cite{Olshanii}. Then, we investigate Casimir force between two
slabs at finite temperature and consider both quantum fluctuation
and thermal fluctuation. Finally, we propose some possible
experimental realization and detection of the Casimir effect. In
this proposal, the geometry of the trapping provides a way to tune
the Casimir force by varying the effective interacting strength
between atoms.

To modify the three dimensional dilute Bose gas into quasi one
dimension, one could apply an axially symmetric two dimensional
harmonic potential to the gas. The effective Hamiltonian of the
trapped interacting Bose gas with s-wave approximation can be
written as
\begin{eqnarray}
\mathscr{H}_{eff}=&-&\frac{1}{2m}\sum_{j=1}^{N}\nabla_{j}^{2}+g\sum_{i<j}\delta(\overrightarrow{r_{i}}-\overrightarrow{r_{j}})\frac{\partial}{\partial
r_{ij}}r_{ij}\nonumber\\
&+&\sum_{j}\frac{1}{2}m\omega_{\bot}^{2}r_{j\bot}^{2},
\end{eqnarray}
where $m$ is the mass of atom, $g$ is the interacting strength
between two atoms, $r_{j\bot}$ is the length of radial position
vector of atom $j$ and $\omega_{\bot}$ is the frequency of the
axially harmonic potential. If the strength of the potential
$\omega_{\bot}$ is large enough, the degrees of freedom of atoms
along axial direction will be frozen so that the atoms can be
considered as moving in one dimensional space. However, the harmonic
potential will renormalize the two-body interacting strength $g$
when you modify the above three-dimensional Hamiltonian into an
effective one-dimensional Hamiltanian
\begin{eqnarray}
\mathscr{H}_{eff}^{1D}=-\frac{1}{2m}\sum_{j=1}^{N}\nabla_{j}^{2}+g_{1D}\sum_{i<j}\delta(x_{i}-x_{j})\frac{\partial}{\partial
x_{ij}}x_{ij},
\end{eqnarray}
where the renormalized interacting strength $g_{1D}$ is
\cite{Olshanii}
\begin{eqnarray}
g_{1D}=\frac{g}{\pi a_{\bot}^{2}}(1-C\frac{a}{a_{\bot}})^{-1},
\end{eqnarray}
where $a_{\bot}=(2\hbar/m\omega_{\bot})^2$ is the characteristic
length of the harmonic potential and $a=gm/4\pi\hbar^2$ is the
s-wave scattering length.
$C=lim_{s\rightarrow\infty}(\int_{0}^{s}\frac{ds'}{\sqrt{s'}}-\sum_{s'=1}^{s}\frac{1}{\sqrt{s'}})=1.4603...
 $ is a constant. This is the Hamiltonian in our further
calculation and it is correct only if the following condition
$\frac{\hbar^{2}p_{z}^{2}}{2m}\ll2\hbar\omega_{\bot}$ is satisfied.
It means that the longitudinal kinetic energy is so small that the
transitions to the excited energy level of the harmonic potential
are safely neglected. In the calculation of ground state energy, we
will take the periodic boundary condition $\varphi(0)=\varphi(d)$
for the wave function, where $d$ is the distance between two slabs.
This condition does not fit the real case, however, it is a good
approximation and convenient for analytical calculation. Through
equation (3), we can see that the effective interacting strength
$g_{1D}$ can be tuned either by varying the depth of the harmonic
potential or by changing the original interacting strength $g$
through a Feshbach resonance. It should be noted that $g_{1D}$ could
be tuned to infinite large when
$(1-C\frac{a}{a_{\bot}})\rightarrow0$. In this case, the system can
be mapped to a noninteracting Fermion gas which is known as
Tonks-Girardeau gas \cite{Tonks}.

\begin{figure}[t]
\includegraphics[
height=2.3523in, width=3.1194in ]{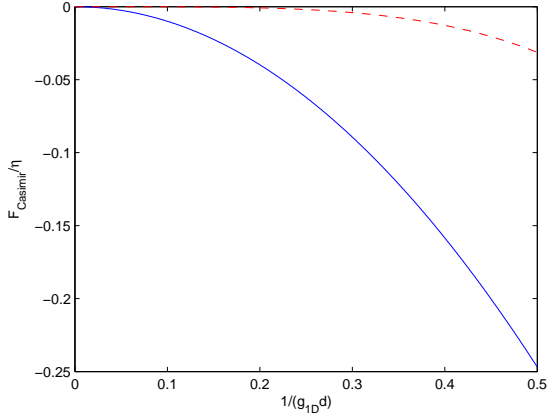} \caption{\label{fig:epsart}
(Color online) The dependence of Casimir force on system length $d$
when $g_{1D}$ is fixed, where $\eta\equiv\pi\sqrt{2\rho}
g_{1D}^{5/2}/12$ is the scale of force. The dash (red) line shows
the Casimir force of a three dimensional interacting Bose gas
without trap calculated in Ref. \cite{Roberts}. The solid (blue)
line shows the Casimir force of Bose gas in a strong harmonic trap
as in equation (8).}
\end{figure}

If we restrict our system to be dilute and weakly interacting, the
ground state energy can be calculated by the standard Bogoliubov
method. The ground state energy of one dimensional Bose gas can be
written as \cite{Lieb}
\begin{eqnarray}
E_{0}=\frac{N^{2}g_{1D}}{d}+\frac{1}{2}\sum_{p\neq0}[\epsilon(p)-p^{2}-2\rho^{2}\gamma],
\end{eqnarray}
where $\epsilon(p)=\rho^{2}p\sqrt{p^{2}+4\gamma}$ is the elemental
excitation spectrum of the quasi-one dimensional interacting Bose
gas, $\rho=N/d$ is the number density of particles and $\gamma\equiv
g_{1D}/2\rho$. It should be noted that $\epsilon(p)$ will give a
phonon type excitation in the long wave-length limit which is very
important to determine the low temperature property of the system.
It is clear that the first term in equation (4) will just give a
zero temperature classical pressure induced by the interaction
between atoms which means that it has no contribution to the Casmir
force. With the boundary condition $ \varphi(0)=\varphi(d)$, the
integral over $p$ will be replaced by a discrete summation. To
calculate the Casimir force, we shall define the Casimir energy as
the difference between the continuum form and the discrete form of
the ground state energy
\begin{eqnarray}
E_{Casimir}=\sum_{n=1}^{+\infty}f(n)-\int_{0}^{+\infty}f(n)dn+\frac{1}{2}f(0),
\end{eqnarray}
where $f(n)\equiv\varrho^{2}[\lambda
n\sqrt{\lambda^{2}n^{2}+4\gamma}-\lambda^{2}n^{2}-2\gamma]$,
$\lambda\equiv\!\pi/(\rho d)$ is just the discrete form of the
second term in equation (4). The physical meaning of this definition
embodies the contribution of the quantum fluctuation at zero
temperature. By using the standard Euler-MacLaurin theorem
\cite{Bordag}, we can write the Casimir energy as

\begin{figure}[t]
\includegraphics[
height=2.3523in, width=3.1194in ]{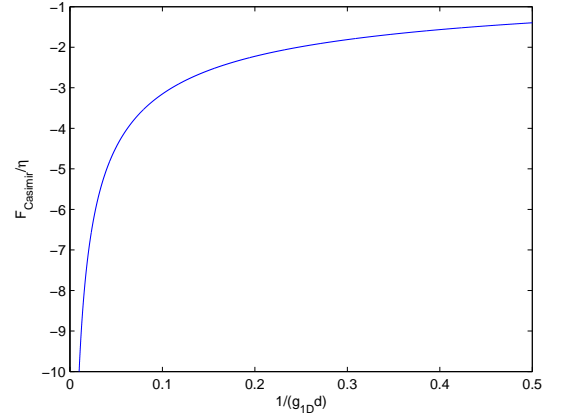} \caption{\label{fig:epsart}
(Color online) The dependence of Casimir force on interacting
strength $g_{1D}$ when length $d$ is fixed which has been shown in
equation (8), where $\eta\equiv\pi\sqrt{2\rho}/(12 d^{5/2})$ is the
scale of force.}
\end{figure}

\begin{eqnarray}
E_{Casimir}=\sum_{n=1}^{\infty}(-1)^{n}\frac{B_{2n}}{(2n)!}f^{2n-1}(0),
\end{eqnarray}
where $B_{2n}$ and $f^{2n-1}(0)$ , $n=1,2,3...$, are the Bernoulli
numbers and the corresponding derivatives at zero point. By
calculating equation (5) directly, we get the final result of the
Casimir energy
\begin{eqnarray}
E_{Casimir}\!=\!-\frac{\rho\pi\sqrt{\gamma}}{6}[\frac{1}{d}-\frac{1}{80g_{1D}d^{2}}+\mathscr{O}(d^{-3})],
\end{eqnarray}
and the Casimir force is
\begin{eqnarray}
F_{Casimir}\!=\!-\frac{\partial E_{Casimir}}{\partial
d}\!=\!&-&\frac{\rho\pi\sqrt{\gamma}}{6d^{2}}+\frac{\rho\pi\sqrt{\gamma}}{240g_{1D}d^{3}}\nonumber\\
&+&\mathscr{O}(d^{-4}).
\end{eqnarray}

we can see that the leading term of $F_{Casimir}$ is proportional to
$d^{-2}$ which is sharply different with the case of three
dimensional geometry \cite{Roberts}, in which the Casimir force is
proportional to $d^{-4}$ (see Fig. 1). It can be seen in Fig 1 that
the Casimir force in three dimensional geomety decay much more
rapidly than the quasi-one dimensional case as the system length
increases. It means that the Casimir effect in one dimension may be
easier to observe in experiments.

Through equation (8) and the expression of $g_{1D}$, it can be seen
that it is possible to vary the Casimir force by tuning the the
characteristic length of the harmonic potential $a_{\bot}$. We can
also vary the interacting strength through Feshbach resonances which
can be easily realized in experiments (see Fig. 2). We can see that
the Casimir force increases rapidly when the interacting strength
increases which is very reasonable because the interaction is the
origin of quasi-particle fluctuation.

\begin{figure}[t]
\includegraphics[
height=2.3523in, width=3.1194in ]{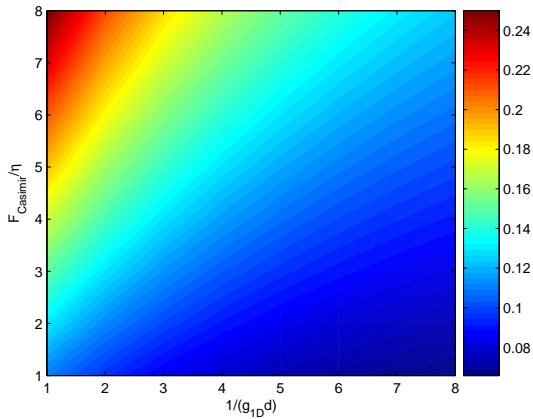} \caption{\label{fig:epsart}
(Color online) The dependence on length $d$ and temperature $T$ of
the Casimir force when $g_{1D}$ is fixed, where $\eta\equiv
g_{1D}^{2}\upsilon_{s}\hbar$ is the scale of force, the temperature
in the color bar is in unit of $g_{1D}\upsilon_{s}\hbar$. The color
shows the temperature dependence. Blue (minimal temperature) and red
(maximal temperature).}
\end{figure}

Furthermore, we will consider the contribution of thermal
fluctuation to the Casimir effect. At finite temperature, the
definition of Casimir force is generalized as \cite{Shyamal1}
\begin{eqnarray}
F_{Casimir}=-\frac{\partial}{\partial
d}[\Omega_{T}(d)-\Omega_{T}(\infty)],
\end{eqnarray}
where $\Omega_{T}$ is the ground potential of our system at
temperature T. In further calculation, we assume that the
temperature is very low so that the quasi-particle energy spectrum
can be approximated as $\varepsilon(p)=\upsilon_{s}p$ like a phonon,
where $\upsilon_{s}=2\rho^{2}\sqrt{\gamma}$ is the speed of phonon.
With this approximation, the system can be considered as a
noninteracting phonon gas, therefore, the ground potential can be
written as \cite{Pitaevskii}
\begin{eqnarray}
\Omega_{T}=-\frac{1}{\beta}\sum_{p}\ln\{1-\exp[-\beta\epsilon(p)]\},
\end{eqnarray}
where the form of $\epsilon(p)$ have been given above. If we take
the boundary condition $ \varphi(0)=\varphi(d)$ into consideration,
the sum over $p$ in equation (9) will have the discrete form
\begin{eqnarray}
\Omega_{T}(d)=-\frac{2}{\beta}\sum_{n=1}^{+\infty}\ln[1-\exp(-\lambda
n)],
\end{eqnarray}
where $\lambda\equiv\frac{2\beta
v_{s}\hbar}{d}=\frac{2\lambda_{T}}{d}$ and $\lambda_{T}$ is the
thermal wave length of the phonon. In the thermodynamic limit,
equation (9) will become a integral form
\begin{eqnarray}
\Omega_{T}(\infty)=-\frac{2}{\beta}\int_{0}^{+\infty}\ln[1-\exp(-\lambda
n)]dn=\frac{\pi^{2}}{3\lambda\beta}.
\end{eqnarray}
Analogous to the case of zero temperature, we also define the
Casimir grand potential as

\begin{figure}[t]
\includegraphics[
height=2.3523in, width=3.1194in ]{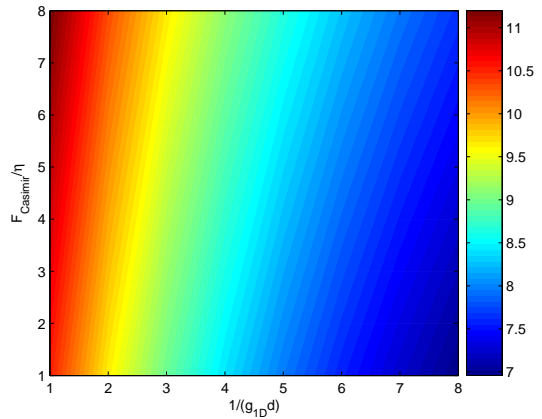} \caption{\label{fig:epsart}
(Color online) The dependence on interacting strength $g_{1D}$ and
temperature $T$ of the Casimir force when length $d$ is fixed, where
$\eta\equiv \sqrt{2\rho}\rho\hbar/(6d^{\frac{5}{2}})$ is the scale
of force, the temperature in the color bar is in unit of $\eta\equiv
\sqrt{2\rho}\rho\hbar/(6d^{\frac{3}{2}})$. The color shows the
temperature dependence. Blue (minimal temperature) and red (maximal
temperature).}
\end{figure}

\begin{eqnarray}
\Omega_{Casimir}\equiv\Omega_{T}(d)-\Omega_{T}(\infty),
\end{eqnarray}
and the Casimir force at finite temperature is
\begin{eqnarray}
F_{Casimir}=-\frac{\partial}{\partial d}\Omega_{Casimir}.
\end{eqnarray}
In our calculation, we just consider the limit $\lambda\ll1$. If the
length of system $d$ is much larger than the thermal wave length
$\lambda_{T}$ (which is mostly true under the real experimental
conditions), this limit will always be satisfied. In this condition,
we can expend the Casimir grand potential in powers of $\lambda$
\begin{eqnarray}
\Omega_{Casimir}=\frac{2}{\beta}[\frac{1}{2}\ln\lambda-\frac{\lambda}{24}+\mathscr{O}(\lambda^{3})]+\mathcal
{K}_{0},
\end{eqnarray}
where $\mathcal
{K}_{0}=\frac{2}{\beta}(\sum_{n=2}^{+\infty}n!a_{n+2}-\frac{11}{12})$
is just a constant and has no contribution to the Casimir force.
Then, the Casimir force can be derived as
\begin{eqnarray}
F_{Casimir}&=&-\frac{\partial}{\partial
d}\Omega_{Casimir}\nonumber\\&=&\frac{1}{d}[k_{B}T-\frac{v_{s}\hbar}{6d}+\mathscr{O}(\frac{1}{d^{3}})].
\end{eqnarray}
It is very interesting that the leading term of this force is
positive which is contrast with the case of three dimensions
 \cite{Lev}. We should also note that the method used to derive Casimir force at finite temperature is fundamentally different from the method in zero temperature, therefore the result of finite temperature cannot go back to the zero temperature solution when $T\rightarrow0$. The dependence on length $d$, interacting strength $g_{1D}$ and
temperature $T$ of the Casimir force is shown in Fig. 3 and Fig. 4.
We can see that at fixed system length $d$ or interacting strength
$g_{1D}$, the Casimir force increases with the temperature because
more quasi-particles(phonons) will be excited by thermal fluctuation
and contribute to the Casimir grand potential. We can also see that
the Casimir force increases when the interacting strength $g_{1D}$
increases which means we can vary the force by tuning $g_{1D}$
through a Feshbach resonance.

To observe the Casimir effect described above, we consider an
atomic-quantum-dot like configuration \cite{Recati,Klein}, which
consists of single impurity atoms confined in a tight trap. In this
configuration, the Casimir effect is represented as the interaction
between the two trapped impurity atoms which can been seen as the
boundary of the quasi-one dimensional bose gas
\cite{Moritz,Recati2}. The interacting energy can be measured by
spectroscopy of a single trapped impurity atom as a function of the
distance between the two impurity atoms. For a quantitative estimate
of this effect, we compute the Casimir grand potential for typical
experimental situations. For typical experimental consideration, the
temperature can as low as 100 $nK$ and length of system is of order
1 $\mu m$, the sound velocity is of order 1 $cm/s$. Under these
conditions, the Casimir grand potential in equation (15) is of order
100 $Hz$ which is experimentally accessible.

In conclusion, we have derived the formula of Casimir force of
quasi-one dimensional Bose gas at zero and finite temperature. The
results show that the Casimir force is very sensitive to the
effective interacting strength which is related to the strength of
the harmonic trapping potential. Another important point we found is
that the Casimir force at finite temperature is positive which is
opposite to the result of three dimensional case in earlier
theoretical work \cite{Lev}. The reason why this happens may be
connected with the dimensional reduction effect in the condensed
matter system. We also propose for the first time an experiment to
control the Casimir force by tuning the frequency of the trapping
potential which has not been considered in earlier experiments.
\\

We express our appreciation for useful discussion with F. Zhou and
S.Q. Shen. This work was supported in part by the project of
knowledge innovation program (PKIP) of Chinese Academy of Sciences,
by NSF of China under grant 10610335, 90406017, 60525417, 10574163,
90306016, the NKBRSF of China under Grant 2005CB724508 and
2006CB921400.

\end{document}